\documentclass{WileyMSP-template}
\usepackage{graphicx}
\usepackage{ragged2e}
\usepackage{lineno}
\usepackage{amsmath}
\usepackage{geometry}
\begin{document}

\pagestyle{fancy}
\rhead{\includegraphics[width=2.5cm]{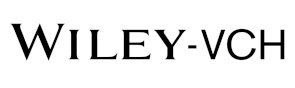}}

\title{Spin Pumping in Magnetostrictive Galfenol Interfaced with Ta}

\maketitle

\author{Ajit Kumar Sahoo$^\dag$},
\author{Suchetana Mukhopadhyay$^\dag$},
\author{Bikram Baghira},
\author{Jeyaramane Arout Chelvane},
\author{Jyoti Ranjan Mohanty},
\author{Anjan Barman*}

\begin{affiliations}
Ajit Kumar Sahoo, Bikram Baghira\\
Department of Condensed Matter and Materials Physics, S. N. Bose National Center for Basic Sciences, Block-JD, Sector III, Salt Lake, Kolkata, 700106, India\\

Suchetana Mukhopadhyay\\
1 Department of Condensed Matter and Materials Physics, S. N. Bose National Center for Basic Sciences, Block-JD, Sector III, Salt Lake, Kolkata, 700106, India\\
2 Department of Physical Sciences, Indian Institute of Science Education and Research Kolkata, Mohanpur, West Bengal, 741252, India

Jeyaramane Arout Chelvane\\
Defence Metallurgical Research Laboratory, Hyderabad, 500058, India

Jyoti Ranjan Mohanty \\
Nanomagnetism and Microscopy Laboratory, Department of Physics, Indian Institute of Technology Hyderabad, Kandi, Sangareddy, Telangana, 502285, India

Anjan Barman\\
Department of Condensed Matter and Materials Physics, S. N. Bose National Center for Basic Sciences, Block-JD, Sector III, Salt Lake, Kolkata, 700106, India\\
$^{\ast}$Email: abarman@bose.res.in\\ 
$^\dag$ Authors with equal contributions.
\end{affiliations}

\keywords{Ultrafast magnetism, Galfenol, Time-resolved magneto-optical Kerr magnetometry, Spin pumping}

\begin{abstract}
In view of their advantages for memory and storage applications, the quest to find suitable magnetic thin film heterostructures that can exhibit strong spin pumping effect persists in the scientific community. Here, the spin pumping phenomenon is investigated in Ta/Galfenol (FeGa) thin film heterostructures by systematically varying the thickness of the heavy metallic Ta underlayer (UL). The films exhibit soft magnetic property with a bcc-phase with a notably low Gilbert damping obtained for FeGa on Si (100). The precessional magnetization dynamics of Ta/FeGa films are explored  using time-resolved magneto-optical Kerr effect magnetometry, revealing the presence of a resonant Kittel mode and additional strain-induced modes. The lowest value of effective Gilbert damping in Ta/FeGa is obtained as $\sim$ 0.015, which rises by $\sim$65\% as the thickness of UL increases. Spin pumping and two-magnon scattering mechanisms are validated using a ballistic spin transport model. An overall effective spin mixing conductance value of $\sim$5.48 $\times$ $10^{15}$ cm$^{-2}$ is found, which is the highest value ever reported in magnetostrictive Galfenol films. Additionally, micromagnetic simulations are performed to understand the effect of tilted magnetic anisotropy on the formation of magnetic modes in these films. These findings in FeGa films establish it as an effective spin source material and offer innovative ideas to control spin-wave propagation and diverse applications in straintronics.
\end{abstract}

\section{Introduction}\label{sec1}
Galfenol (FeGa) has attracted significant attention in the scientific community due to its remarkable mechanical properties, low coercive field value, and magnetostrictive behavior \cite{1}-\cite{3}. FeGa exhibits high magnetostriction i.e, magnetization changes substantially in response to mechanical stress or applied magnetic field \cite{4}-\cite{6}. These distinctive features of FeGa have placed it as a versatile material for various applications, e.g, in magnetic sensors, actuators, and storage technologies like acoustically assisted magnetic recording (AAMR) \cite{7}-\cite{9}. In such AAMR technology, a surface acoustic wave (SAW) is used to manipulate the coercivity of the recording medium through inverse magnetostriction. FeGa films can also be used in applications requiring reduced hysteresis loss and minimized heat dissipation in devices with repeated magnetization changes, such as integrated circuits \cite{10}. FeGa and its derivative amorphous alloy FeGaB are also well known for their soft magnetic properties \cite{10a, 10b}. Tuned magnetic anisotropy and Gilbert damping are possible in Al$_2$O$_3$/FeGaB thin films and therefore Ta/FeGaB/Al$_2$O$_3$-based heterostructures \cite{10c, 10d} have been proposed as candidates for microwave and frequency tuning applications. The notable properties of these films arise from its intricate microstructure and crystallographic nature \cite{11}-\cite{13}. Controlling their magnetic properties is crucial for tuning them to some specific applications. For instance, the structural, magnetic, and magnetostrictive properties of the FeGa systems can be tuned by introducing a non-magnetic underlayer (UL) or rare-earth materials \cite{3}, \cite{5}, \cite{11}, \cite{14}.   

\vspace{7pt}
 
FeGa films can exhibit both in-plane and perpendicular magnetic anisotropy, which is crucial for diverse applications in spintronic devices \cite{12}, \cite{15}, \cite{15-a}. Magnetic domains in FeGa/BaTiO$_3$ bilayers, which show in-plane magnetic anisotropy, can be controlled using static voltage through electromechanical coupling \cite{16}. However, the concept of introducing tilted magnetic anisotropy in thin films has also emerged as an area of significant importance. By deliberately engineering the magnetic anisotropy axis from the plane of the film, the magnetic properties and magnetization dynamics can be tuned as per the desired requirements \cite{17}-\cite{20}. It is important to note that, introduction of an UL can also influence the anisotropy axis of FeGa films. This additional UL, often comprising a different material, can significantly influence the growth and structural arrangement of  the overlying FeGa film \cite{5}, \cite{13}. In such cases, the UL acts as a seed layer that can promote improved crystallinity, reduced defects, and enhanced magnetic response within the thin film. It has been reported that the presence of a few nanometers of Ta, Cu, and NiFe ULs can significantly decrease coercivity and the effective Gilbert damping \cite{13}, \cite{21}. Importantly, the UL can also minimize lattice strain that typically arises in thin films due to lattice mismatch with the substrate that can detrimentally affect its magnetic damping.

\vspace{7pt}

Another intriguing phenomenon that can arise from the interface of FeGa film and UL is spin pumping, which involves the precessing magnetization in a ferromagnetic layer inducing a spin current flow into an adjacent layer \cite{18, 22}. The manipulation of pure spin current in the simplest heterostructure system consisting of a ferromagnet (FM) and a nonmagnet (NM) layer involves three critical aspects, namely, its generation in the FM, transport across the interface, and absorption in the NM \cite{22-a}. Various physical phenomena, including the spin Hall effect (SHE) \cite{22-b}, spin caloric effects \cite{22c}, \cite{22d}, spin pumping \cite{22e}, and non-local injection of spin \cite{22f}, have been harnessed for the efficient generation of pure spin current in NM/FM heterostructures. Among these, spin pumping has attracted widespread interest due to the simplicity of device structure and its potential in spintronics applications \cite{23}-\cite{25}. Various methods have been employed to investigate spin pumping, including ferromagnetic resonance (FMR) and inverse spin Hall effect (ISHE) measurements \cite{26}-\cite{29}. However, our study leverages the capabilities of all-optical time-resolved magneto-optical Kerr effect (TR-MOKE) measurements to explore the spin pumping effect in Ta/FeGa thin films with tilted magnetic anisotropy. Recently, magnetic damping and ultrafast demagnetization have been reported in FeGa and FeGa-based bilayer thin films \cite{30}-\cite{32}. Anisotropic magnetoelastic Gilbert damping has also been observed in FeGa thin films over a range of temperatures \cite{33}. One of the main research gaps in the Galfenol system is the understanding of the interplay between tilted magnetic anisotropy, strain effects, and ultrafast magnetization dynamics. Extensive research is required to outline how these properties can be controlled to achieve the desired magnetization dynamics in Galfenol films. Here, we attempt to close this gap by carrying out a systematic investigation of magnetization dynamics and spin pumping in Ta/FeGa bilayer films at room temperature (RT).  

\vspace{7pt}

In this study, we investigate the structural and magnetic properties of Si/Ta($t_{NM}$)/FeGa($t_{FM}$)/Ta(2 nm) magnetic films. $t_{NM}$ and $t_{FM}$ denote thickness of Ta UL, and FeGa layer respectively. Heavy metallic Ta can exhibit good spin-sinking properties which makes it effective for spin-based electronics devices. To study the spin pumping as a function of Ta thickness, we systematically varied $t_{NM}$ from 1 nm to 15 nm. Structural investigations reveal that all the Ta/FeGa films are crystalline in nature. All films show low magnetic coercivity and a uniform angular variation of the in-plane coercive field. Tilted magnetic anisotropy and weak contrast in magnetic domains are observed. We use TR-MOKE magnetometry to study the magnetization dynamics and the ensuing spin pumping phenomenon in these samples and extract the spin-mixing conductances and the coefficient of two-magnon scattering (TMS) at the Ta/FeGa interface. Finally, we qualitatively validate our experimental understanding of the effect of tilted magnetic anisotropy on the magnetization dynamics of FeGa films using micromagnetic framework.   
 
\section{Results and Discussion}\label{sec2}

\begin{figure} 
\centerline{\includegraphics[scale=0.7]{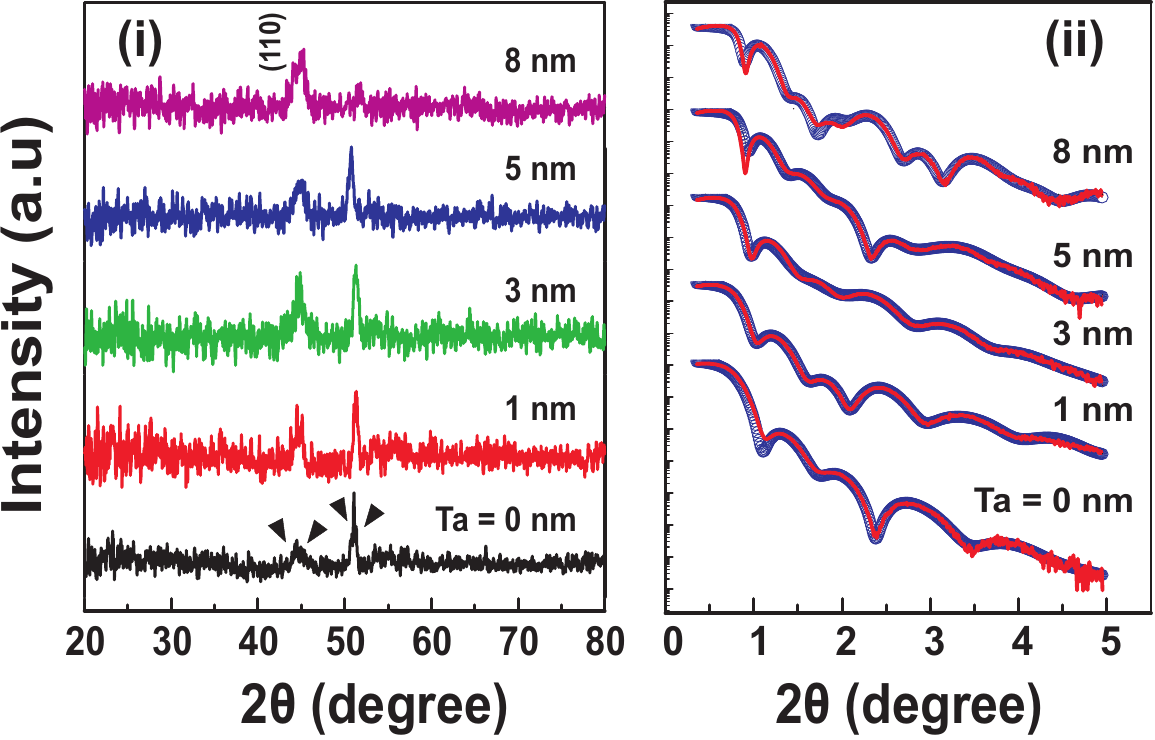}}

\caption{Analysis of the structure and thickness of thin films involving two main aspects: (i) X-ray diffraction patterns of Ta(t)/ FeGa samples with varying thickness of Ta UL ($t_{NM}$ = 0, 1, 3, 5, and 8 nm) and (ii) X-ray reflectivity (XRR) spectra of the corresponding samples. In this representation, the red curves represent the experimental data, while the blue curves depict the theoretical fits where (a. u.) stands for arbitrary units. }\label{F1}
\end{figure}

\begin{figure} 
\centerline{\includegraphics[scale=0.5]{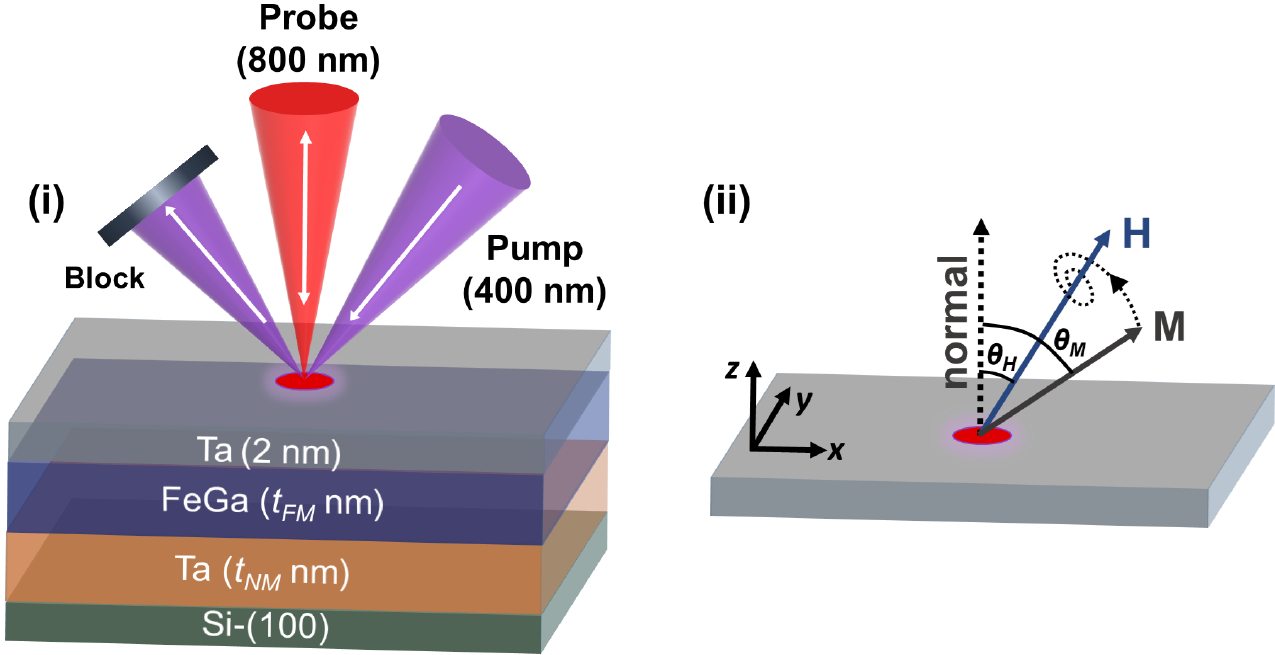}}
\caption{A sketch of the sample stack with the TR-MOKE pump and probe geometry is shown. (i) Si (100) is used as a substrate, and Ta of various thicknesses are inserted as an underlayer in FeGa film. Ta (2 nm) is used as a capping layer to prevent sample degradation. The pump (wavelength: 400 nm) and probe (wavelength: 800 nm) beams are incident on the sample in a non-collinear geometry. The spot size of the pump (blue) is larger than that of the probe (red). (ii) depicts the precession of a typical effective magnetization ($\textbf{M}$) of the sample around the effective field ($\textbf{H}$). Three axes ($\textbf{X}$, $\textbf{Y}$, $\textbf{Z}$) are mentioned with reference to the plane of the sample. Whereas, $\theta_{H}$ and $\theta_{M}$ denote the angles of orientation of $\textbf{H}$ and $\textbf{M}$ from the normal orientation. }\label{F2}
\end{figure}

FeGa thin films are deposited either onto a bare Si substrate or on a Ta layer deposited on an Si substrate (see Experimental Section). Figure \ref{F1}(i) shows the grazing incidence X-ray diffraction (GIXRD) patterns, taken at an angle of $0.5^{\circ}$, of Ta($t_{NM}$)/FeGa samples with varying thickness ($t_{NM}$ = 0, 1, 3, 5, and 8 nm) of the Ta UL. A sketch of the sample stack is presented in Figure \ref{F2}(i). The XRD peaks identify the presence of FeGa, Ta ($\beta$-phase), and FeTa in the sample. Peaks are observed at two distinct positions, around 2$\theta$ = $44.50^{\circ}$ and $51.08^{\circ}$. Each peak position contains three different peaks (a zoomed version is shown in Figure S1 of the Supporting Information). Around 2$\theta$ = $44.50^{\circ}$ of Ta (312), two other distinct peaks are observed at $43.94^{\circ}$ and $45.02^{\circ}$ of FeGa (110) and Ta (510), respectively. Similarly, three different peaks are observed at $50.78^{\circ}$, $51.08^{\circ}$, and $51.32^{\circ}$ of Ta and FeTa phase formation in the sample. The FeGa (110) peak indicates the formation of the bcc-phase in the film, as previously reported \cite{12}. In addition, XRD peak splitting occurs due to various factors such as lattice strain, structural defects, phase transitions, etc. \cite{34}, of which lattice strain is the most probable reason for the splitting observed in our system. The lattice constants of Si, Ta, and FeGa are 5.431, 3.305, and 2.933 \AA, respectively. The lattice mismatch between the Si substrate and the FeGa film is calculated as 45\%, whereas the lattice mismatch is reduced to 11\% when Ta is inserted between the substrate and the FeGa film. Thus, the presence of the Ta UL significantly reduces lattice strain in the FeGa film. The small peak shift due to the relative change in strain, can be understood using Bragg's law as: $\Delta \epsilon$ = $sin\theta_2$/$sin\theta_1$ - 1 \cite{21}. To obtain more information about the film thickness and interfacial roughness, we performed XRR analysis for $t_{NM}$ = 0, 1, 3, 5, and 8 nm. Figure \ref{F1}(ii) represents the Kiessig fringes of all these samples. The experimental XRR spectra were analyzed using GenX software, employing a specular simulation with the interdiffusion model \cite{34-a}. We found that the observed film thickness is comparable to the nominal film thickness (see Table S1 of the Supporting Information). Moreover, an interfacial layer of FeTa is formed, and the roughness of the interfacial layer increases as $t_{NM}$ increases.

\vspace{7pt}

\vspace{7pt}

\begin{figure} 
\centerline{\includegraphics[scale=0.8]{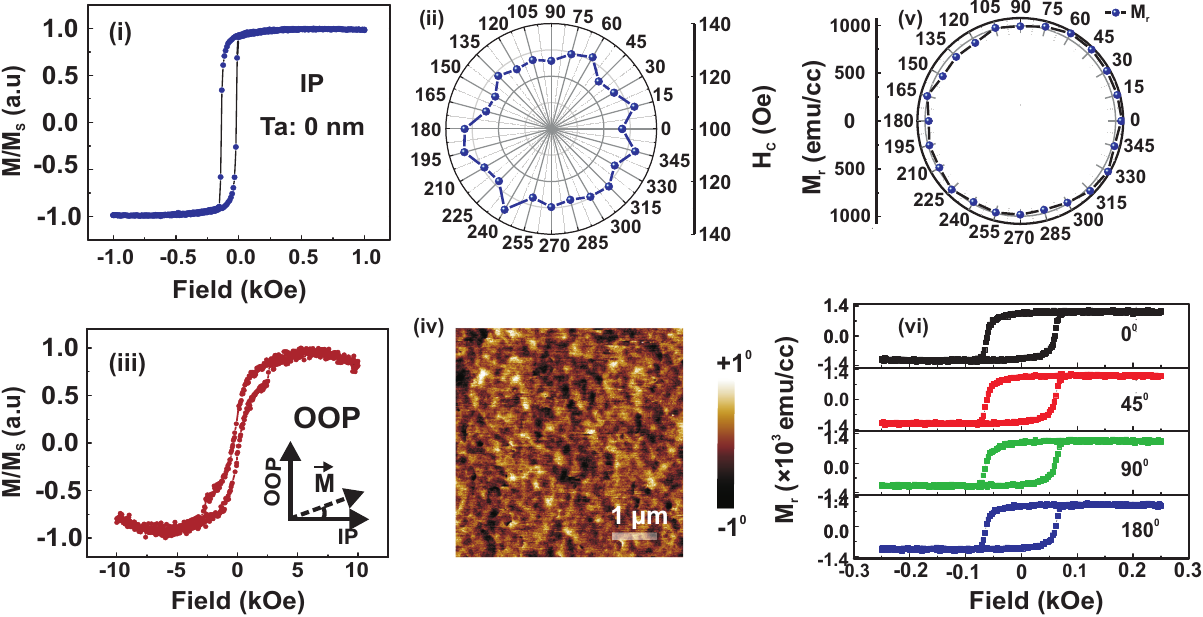}}
\caption{Static magnetic properties and magnetic domain analysis of the FeGa sample, where $t_{NM}$ = 0 nm, are presented in this section. (i) The hysteresis loop is shown along the  IP direction, (ii) a polar plot displays the variation of coercive field ($H_{C}$) from $0^{0}$ to $360^{0}$ in the IP configuration, (iii) the hysteresis loop along the OOP direction is presented, with the inset illustrating the effective magnetization ($\vec{M}$) oriented more towards the IP direction, and, (iv) the Magnetic Force Microscopy (MFM) image of the sample with Ta UL = 0 nm, (v) $M_r$ values are shown in a polar plot display, (vi) the representative hysteresis loops are shown at 0, 45, 90, and $180\degree$.}\label{F3}
\end{figure}

Next, we investigated the static magnetic properties of FeGa films in the presence of a Ta UL. Static magnetic characterization is presented in Figure \ref{F3}. The effects of the Ta non-magnetic UL on static magnetic properties, such as coercive field ($H_C$), saturation magnetization ($M_S$), and remanent magnetization ($M_r$), are found to be minimal. Therefore, we present the hysteresis loops for 7 nm-thick FeGa films without a Ta UL (Ta = 0 nm). We have not observed any significant change in $H_C$ and $M_S$ in the presence of the Ta UL (see Table S2 of the Supporting Information). A similar observation of unchanged $M_S$ values was also reported in CoFeB thin films with Ta UL thickness varying from 0 to 20 nm \cite{22}. Figure \ref{F3}(i) shows the in-plane (IP) hysteresis of Ta = 0 nm sample. The slight shift of the hysteresis loop towards negative fields in Figure 3 (i) is due to the small remnant field present in the superconducting magnet used of the PPMS even after degaussing. However, this does not affect the extraction of magnetic parameters like the $H_C$ and $M_S$ values. IP $H_C$ and $M_S$ are calculated as 80 $\pm$ 10 Oe and 1200 $\pm$ 200 emu/cc, respectively. To understand the nature of magnetic anisotropy in the film, we studied the angular dependence of $H_C$ along the IP direction of Ta = 0 nm, as shown in Figure \ref{F3}(ii). The observed $H_C$ values are nearly equal, within the error margins, along the angular variation of the IP applied field.

\vspace{7pt}

According to previous reports \cite{35}, the angular dependence of the squareness ratio ($M_r$/$M_S$) of the FeGa film reveals no uniaxial symmetry in both IP and hard-axis directions. We performed VSM measurements of 7 nm-thick FeGa film to verify this. As seen from the angular dependence of the remanent magnetization in Figures  \ref{F3} (v) and (vi), the FeGa film does not show any anisotropic nature in the sample plane. However, it does possess an out-of-plane (OOP) magnetization component. Figure \ref{F3}(iii) displays the hysteresis loop of OOP measurements for Ta = 0 nm sample. We observed a small opening ($H_C$ = 255 $\pm$ 10 Oe) and tilting in the OOP hysteresis. The saturation field ($H_S$) is observed to be 0.1 kOe and 3 kOe along the IP and OOP directions respectively. It should be noted that the FeGa film saturates more easily along the IP orientation as compared to the OOP orientation. Therefore, the effective orientation of magnetization lies closer to the IP direction of the film, as shown in the inset of Figure \ref{F3}(iii). However, the effective magnetic anisotropy of the film does not align with either the IP or OOP orientation but exhibits a tilted orientation. Figure \ref{F3}(iv) illustrates the magnetic contrast of the FeGa film using an OOP magnetized probe. Due to the tilted nature of magnetic anisotropy in the film, we observed weak magnetic domain contrast. It is worth noting that Ga-rich Fe$_{72}$Ga$_{28}$ films exhibit an OOP component of magnetization, which promotes magnetic ripples in the film \cite{15-a}. The OOP component of the magnetization is reduced upon slight annealing, resulting in the formation of more IP components of the magnetization. The coexistence of both IP and OOP components of magnetization break the uniform orientation of the magnetization which may form random magnetic anisotropy in FeGa films \cite{37}. The lack of orientation of the magnetization is attributed to various reasons, such as simultaneous presence of distinct crystal phases, interface magnetic anisotropy, and non-uniform dispersion of internal strain in FeGa films.

\vspace{7pt}

\begin{figure} 
\centerline{\includegraphics[scale=0.6]{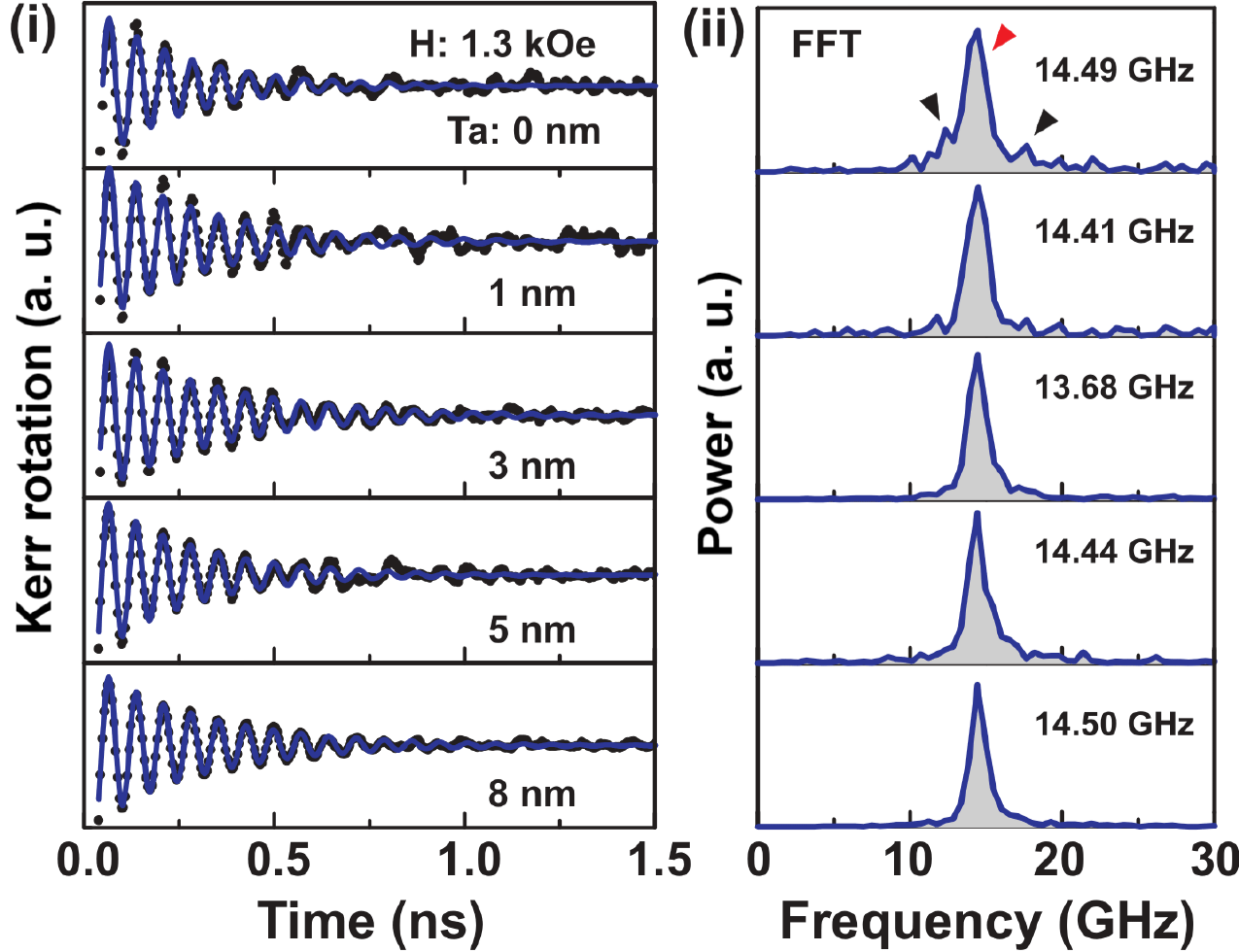}}
\caption{Time-resolved magneto-optical Kerr effect (TR-MOKE) data and corresponding fast Fourier transformed (FFT) power spectra:  (i) Background-subtracted magnetization precession data, depicted as black dots, for FeGa samples with varying thicknesses of Ta underlayer (UL). These data are obtained using TR-MOKE magnetometry and fitted (blue curves) using a phenomenological equation. (ii) The corresponding FFT spectra of the samples are shown, where the dominant Kittel mode is indicated by a red arrow (its value is provided in the inset), and additional strain-induced modes are indicated by black arrows. The details of the modes are discussed in the following sections.}\label{F4}
\end{figure}

The ultrafast magnetization dynamics of the Ta/FeGa films are investigated using TR-MOKE technique. To determine the prominent precessional modes in these films, we analyzed the nanosecond precessional dynamics by applying bias magnetic field of different magnitudes. Figure \ref{F4}(i) represents the background-subtracted time-resolved data for various films, such as Ta = 0, 1, 3, 5, and 8 nm. We conducted a fast Fourier transformation (FFT) of the corresponding precessional data to obtain the spin-wave (SW) spectra in the frequency domain, as shown in Figure \ref{F4}(ii). In the case of FeGa film, i.e., Ta UL = 0 nm, we obtained the most intense FFT peak at a frequency of 14.49 GHz (marked with a red arrow), which is identified as the Kittel mode. Additionally, some other low-intensity modes (marked with black arrows) appeared below and above the Kittel mode. These modes likely arise due to the presence of magnetostriction in the FeGa film. It is reported that strain from the substrate or external source can induce multimode SW excitation in magnetic thin films \cite{38}, \cite{38a}. Depending on the magnitude of external strain and the type of strain, dipolar magnetostatic surface wave (MSSW) modes, and exchange perpendicular standing SW (PSSW) modes can be induced \cite{39}. Since we established that the insertion of a Ta UL between the substrate and the FeGa film reduces lattice strain, intensity of the strain-induced modes become softer in the presence of the UL. In metal-insulator strained epitaxial films with dominant interface anisotropy, Damon-Eshbach (DE) and PSSW modes can also appear \cite{40}. However, DE modes are generally observed in the DE geometry using Brillouin light scattering spectroscopy and are not observed here.

\vspace{7pt}

In our case, the Kittel mode appears roughly around the same frequency for all the Ta/FeGa  films with various UL thicknesses, as shown in Figure \ref{F4}(ii). The films exhibit distinct precession and slowly decay over a time period of 1.5 ns. To determine the decay time, we fit the time variation of precessional oscillations of Kerr rotation using the phenomenological equation, as given below:

\vspace{7pt}

\begin{equation}
\theta_{K} \propto A exp(-t/\tau)sin[2\pi(f+bt)t + \phi_{0}],
\end{equation}
\vspace{5pt}

where $A$, $\tau$, $f$, and $\phi_{0}$ represent the initial amplitude of the magnetization precession, decay or relaxation time, precession frequency, and initial phase, respectively. The parameter $b$ accounts for the frequency shift over time due to pump-induced changes in the magnetic parameters within the probed area, and lies within 4.79 $\pm$ 3.68 $\times$ $10^{17}$ GHzns$^{-1}$ in the considered films. Notably, the critical parameter $\tau$ for the Kittel mode of samples with Ta = 0, 1, 3, 5, and 8 nm is calculated as 0.18, 0.20, 0.22, 0.25, and 0.28 ns, respectively. Comparing Ta = 8 nm to Ta = 0 nm film, we observed the increment in $\tau$ as $\approx$ 55\%. This increment in $\tau$ with the increase in the thickness of Ta UL may be attributed to spin diffusion at the interface between the Ta UL and the FeGa film. Additionally, it is worth mentioning that the relaxation times for the other modes are observed to be approximately 90\% shorter than that of the Kittel mode.

\vspace{7pt}

\vspace{7pt}

To understand the anisotropic behavior of these modes, we measured the time-resolved magnetic response while varying the bias magnetic field from 1.3 kOe to 0.7 kOe. Figure \ref{F5}(i) presents the time-resolved data for the FeGa sample (thickness of Ta UL = 0 nm) with variations of the field at 1.2, 1.1, and 0.9 kOe. The FFT of the corresponding precessional data is shown in Figure \ref{F5}(ii). In addition to the uniform Kittel mode, we observed other modes in the vicinity of the uniform precession mode. The positions of these other modes also changed with the application of the magnetic field identifying them as strain-induced magnetic modes. The shifting of the uniform Kittel mode is indicated by the dashed lines, and the strain-induced modes are marked with black arrows ($f_{K1}$, $f_{K2}$, and $f_{K3}$). This behavior holds true for various thicknesses of the Ta UL in FeGa films, as depicted in Figure S3 (Supporting Information). To understand the nature of the uniform precession mode, we fitted our frequency vs. field data for Ta = 0 nm sample with the Kittel equation:   

\begin{figure} 
\centerline{\includegraphics[scale=0.7]{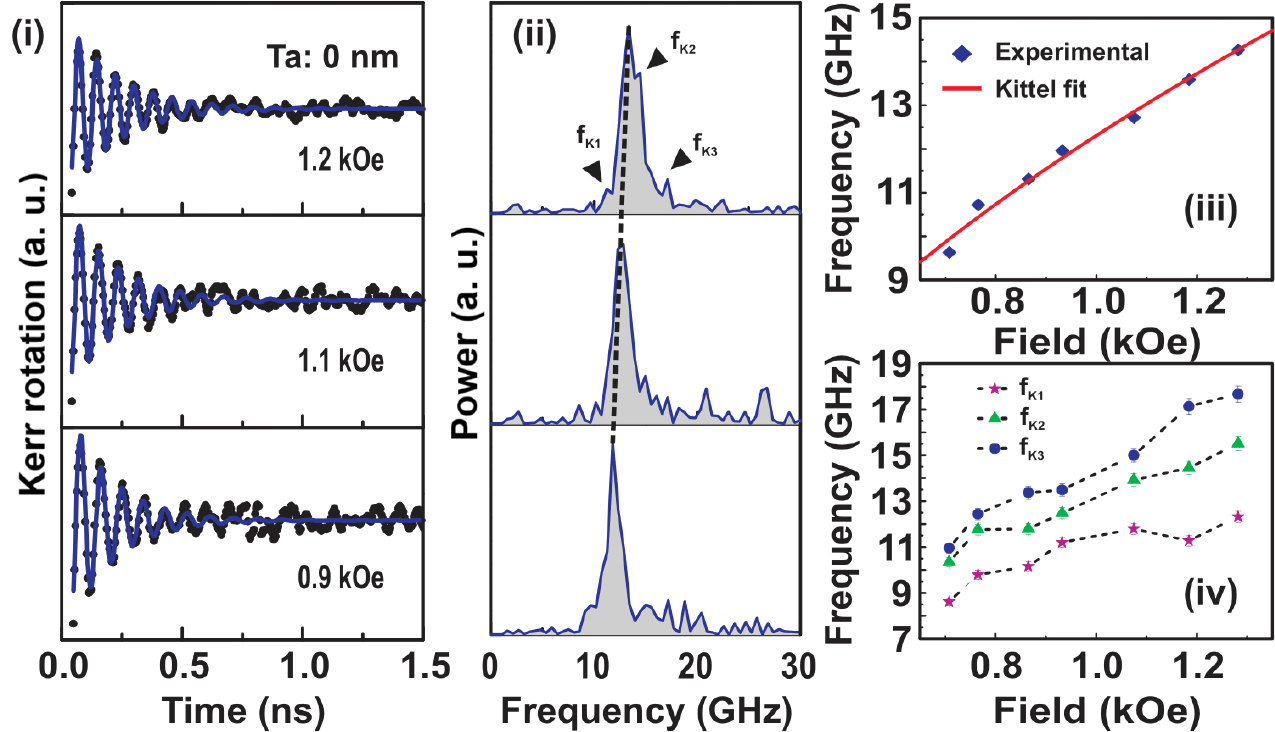}}
\caption{Field-dependent TR-MOKE data of sample Ta UL = 0 nm are presented as follows: (i) Background-subtracted magnetization precession data (blue curves represent fits using a phenomenological equation) at various applied fields (as mentioned in the inset). (ii) Corresponding FFT spectra are shown, with black dotted lines indicating the shift of the Kittel mode, and black arrows representing strain-induced modes other than the Kittel mode. (iii) The bias magnetic field dependence of the precessional frequency is depicted, with a red solid line representing the Kittel fit. (iv) The bias field dependence of the strain-induced modes, e.g, $f_{K1}$, $f_{K2}$, and $f_{K3}$ are presented here, where dotted lines are guides to the eye.}\label{F5}
\end{figure}

\vspace{7pt}
  
\begin{equation}
f=\frac{\gamma}{2\pi} \sqrt{(H_{1} \times H_{2})}
\end{equation}

\begin{equation}
H_{1}=Hcos(\theta_{H} - \theta_{M}) - 4\pi M_{eff}cos^{2}(\theta_{M}),
\end{equation}

\begin{equation}
H_{2}=Hcos(\theta_{H} - \theta_{M}) - 4\pi M_{eff}cos(2\theta_{M}).
\end{equation}

Here, $f$ represents the precessional frequency in the magnetic system, $\gamma$ stands for the gyromagnetic ratio, $H$ corresponds to the magnitude of the bias field, and $\theta_{H}$ and $\theta_{M}$ denote the angles of orientation of $H$ and $M$, respectively, from the normal (as depicted in Figure \ref{F2}(ii)). $M_{eff}$ signifies the effective magnetization, which can be correlated with the saturation magnetization ($M_S$), magnetic anisotropy constant ($K_s$), and ferromagnetic film thickness ($t_{FM}$), as shown below \cite{41}:

\begin{equation}
4\pi M_{eff}=4\pi M_{S}-\big(\frac{2K_{s}}{M_{S}}\big) t_{FM}^{-1}
\end{equation}
\vspace{5pt}

For fitting to the Kittel equation, $\theta_{H}$, $\theta_{M}$, and $M_{eff}$ are taken as the fitting parameters. The fitted values for $\theta_{H}$ and $\theta_{M}$ are $23.4\pm0.5^\circ$ and $67.5\pm0.5^\circ$, respectively. By utilizing Equations 2 and 5, the calculated $M_S$ value is 1212.83 emu/cc, whereas $K_s$ is obtained as -0.53 ergcm$^{-2}$. The calculated value for $M_S$ lies within the range of the value obtained by the physical property measurement system (PPMS) measurement (see Table S2 of the Supporting Information). For $M_S$ = $1300\pm100$ emu/cc, effective anisotropy ($K_{eff}$) is obtained in the range of -$(3.6-5.7)\pm10^5$ ergcm$^{-3}$. The value of $K_{eff}$ can be estimated using the expression: ($\mu_{0}$/2)$\times$($M_S^2$-$M_{eff}^2$). Our calculated anisotropy constant agrees with the previously reported anisotropy constant values of FeGa films \cite{42}. After obtaining the values of $\tau$ and $K_{eff}$ from the above equations, we calculate the effective Gilbert damping parameter ($\alpha_{eff}$) using the following equation:  

\begin{equation}
\alpha_{eff} = \frac{1}{\gamma \tau (H+\frac{2 K_{eff}}{M_{S}}+2 \pi M_{S})}
\end{equation}
\vspace{5pt}

\begin{figure} 
\centerline{\includegraphics[scale=0.6]{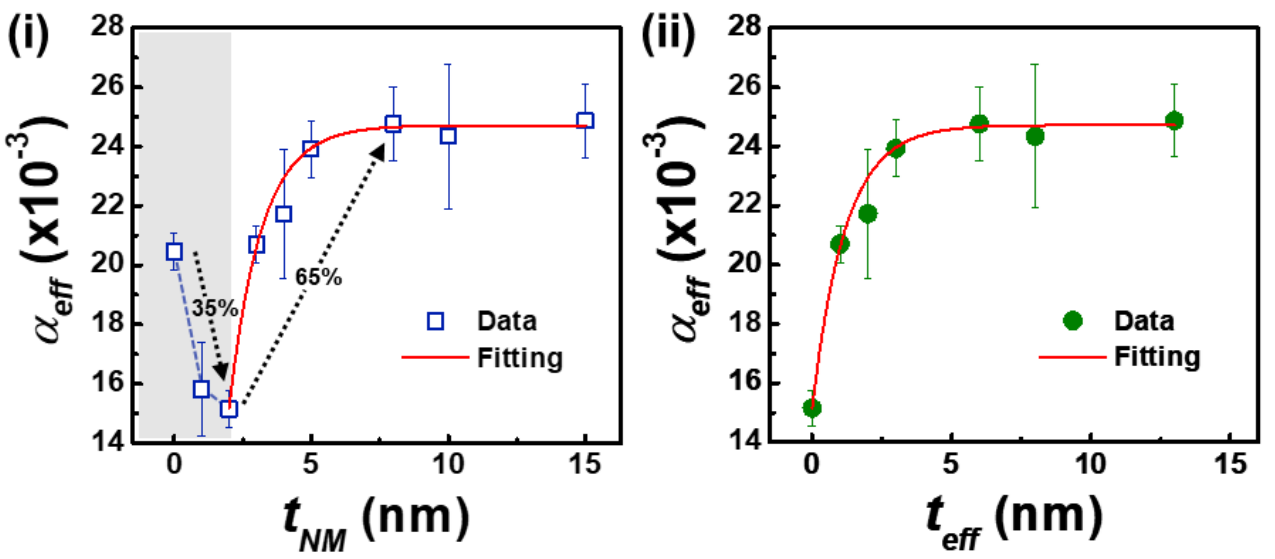}}
\caption{Effective magnetic damping $\alpha_{eff}$ for various thicknesses of the Ta UL. (i) Variation of $\alpha_{eff}$ with the thickness $t_{NM}$ of Ta UL. The gray shaded region corresponds to the regime dominated by the effect of strain, (ii) Fitting with the ballistic spin transport model considering $\alpha_{eff}$ vs $t_{eff}$ = ($t_{NM}$ - 2 nm).}\label{F6}
\end{figure}

The extracted values of $\alpha_{eff}$ for various thicknesses of Ta UL are shown in Figure \ref{F6}. The $\alpha_{eff}$ of 7 nm-thick FeGa film with no Ta UL is $\approx$ 0.0205, which is notably low for FeGa on Si (100) compared to previous reports \cite{21}, which can be attributed to our carefully optimized sputter deposition. The value of $\alpha_{eff}$ of FeGa films, obtained using FMR technique, fall within the range of 1 to 4.5 $\times$ $10^{-2}$ \cite{33}. The $\alpha_{eff}$ in our case decreases by $\approx$ 41\% when Ta UL thickness increases to 2 nm. This reduction arises from the lowering of lattice strain by the insertion of the Ta UL, which causes a reduction in strain relaxation within the film. Increased magnetic damping in the bare FeGa film is also expected due to stronger excitation of multimode spin waves, which introduce additional pathways for energy dissipation, and thereby influence the behavior of the Kittel mode \cite{38}. As the UL thickness increases to 8 nm, $\alpha_{eff}$ increases by $\sim$65\% and saturates at $\approx$ 0.0248. The increment in effective damping with increasing Ta UL thickness beyond 2 nm followed by a characteristic saturation arises due to the spin pumping phenomenon \cite{22e}.

\vspace{7pt} 

In NM/FM heterostructures, spin pumping from the resonant magnetization precession in an FM causes a spin current injection into the interfacing heavy metallic NM layer leading to an increase in the effective Gilbert damping of the FM \cite{22e}. It is a reciprocal effect to the spin-orbit torques (SOTs), wherein passing an electric current in the interfacing NM layer can generate pure spin currents that can in turn exert a torque on the FM magnetization \cite{43}. The SHE originates due to strong spin-orbit coupling in heavy materials and results in the generation of spin-polarized electrons at the material's surface when an electric current is applied to it \cite{43aa}. If the material is interfaced with an FM, it can exert SOT on the precessing magnetization of the FM. The spin Hall SOT can modulate the Gilbert damping \cite{43a}, while the interfacial Rashba effect can additionally contribute to the field-like SOT \cite{43b}. These SOTs can be quantified using the dc-tuned ST-FMR technique \cite{43, 43a} from the change in linewidth (or $\alpha_{eff}$) due to dc bias current $I_{dc}$ (for damping-like SOT) or the change in resonance condition (for field-like SOT). In contrast, in our dynamical spin pumping measurements on Ta/FeGa system, no dc bias current is applied, and the enhancement $\Delta\alpha$ of the Gilbert damping parameter is sensitive to the spin sinking by the heavy metallic Ta underlayer, and other extrinsic sources of spin scattering at the interface. While parasitic spin pumping-induced ISHE voltage can manifest in ST-FMR voltage having opposite sign to its symmetric component \cite{43c}, this contribution is typically quite small \cite{43d, 43e}, and arises due to the resonant magnetization dynamics of the FM being excited. However, residual SOT effects such as SHE of Ta are not anticipated in spin pumping measurements because no dc current is applied to the Ta without which it cannot generate spin currents, unlike a ferromagnetic spin source which generates spin currents as a result of its magnetization precession \cite{22e}. This is routine for spin pumping studies in NM/FM systems wherein no dc bias current is applied and thereby SOT effects can be neglected \cite{43f}-\cite{43h}. 

\vspace{7pt}

The damping at the interface between Ta and FeGa film can be linked to the interfacial spin-mixing conductance of Ta/FeGa system. Two definitions of the spin-mixing conductance are commonly considered, the first of which is the intrinsic spin mixing conductance ($G_{\uparrow\downarrow}$), which straightforwardly characterizes spin current transmission at the interface without accounting for the backflow of spin angular momentum. The second is the effective spin-mixing conductance ($G_{eff}$), which takes into consideration the spin backflow \cite{44, 44b}. Considering the spin transmission at the interface as the sole underlying cause of the enhanced magnetic damping in Ta/FeGa, we can elucidate the connection between these two parameters through the following equation \cite{45}: 
 
\begin{equation}
G_{eff}=G_{\uparrow\downarrow}(1-e^{-\frac{2t_{NM}}{\lambda_{sd}}})
\end{equation}

where $\lambda_{sd}$ denotes the spin diffusion length of Ta. The effective damping modulation in the presence of spin pumping and TMS is given by \cite{45_a}: 

\begin{equation}
\alpha_{eff}=\alpha_{0}+\frac{g\mu_{B}G_{eff}}{4 \pi t_{FM}M_{eff}}+\frac{\beta_{TMS}}{t_{FM}^{2}}\\
\end{equation}
\vspace{5pt}

where $\beta_{TMS}$ is the TMS coefficient. Damping enhancement due to TMS follows a parabolic dependence on the inverse of FM thickness, unlike the inverse linear dependence due to spin pumping. As XRR analysis suggests, a small interfacial layer is formed between the Ta and FeGa layers. TMS is crucial to consider when interfacial roughness and intermixing exist in thin-film multilayers because it contributes to SW scattering at defects or irregularities. This scattering process impacts the propagation of SWs, affecting the overall magnetic behavior, and can be particularly relevant for understanding and controlling SW transport in magnetic systems. Substituting the expression for $G_{eff}$ from Equation 7 and taking $\alpha_{0}^{'}=\alpha_{0}+\beta_{TMS}/t_{FM}^{2}$, we have 

\begin{equation}
\alpha_{eff}=\alpha^{'}_{0}+\frac{g\mu_{B}G_{\uparrow\downarrow}(1-e^{-\frac{2t_{NM}}{\lambda_{sd}}})}{4 \pi t_{FM}M_{eff}}
\end{equation}
\vspace{5pt}

\vspace{5pt}

To extract the intrinsic spin-mixing conductance ($G_{\uparrow\downarrow}$) of Ta/FeGa system and $\lambda_{sd}$ in Ta, we used a ballistic spin transport model as it is conventionally applied (Equation 9) to fit the variation of  $\alpha_{eff}$ versus $t_{eff}$ = ($t_{NM}$ - 2 nm) where $t_{NM}$ is the thickness of the Ta UL as shown in Figure 6 (ii).  $\alpha_{0}^{'}$ is taken as the damping corresponding to Ta UL of 2 nm, beyond which the additional damping due to strain relaxation (present for UL thickness  = 0 and 1 nm) is eliminated. The red line in Figure \ref{F6} represents the best fit for the variation of $\alpha_{eff}$. For our case, the value of $\lambda_{sd}$ is nearly 2.37 nm, which is close to the reported value of 2.44 nm for the Ta/CoFeB system \cite{22}. The value of $G_{\uparrow\downarrow}$ is extracted to be (5.88 $\pm$ 0.08) $\times$ $10^{15}$ cm$^{-2}$. The values of $G_{eff}$ calculated for various UL thickness using Equation 7 are provided in Table S4 of the Supporting Information. It is important to mention that we observed the highest $G_{eff}$ values ever reported in magnetostrictive thin films. The $G_{eff}$ values are comparable to the values of effective spin mixing conductance reported from spin pumping measurements in magnetostrictive ferromagnet-based Fe$_{78}$Ga$_{13}$B$_9$/Ag/Bi$_{85}$Sb$_{15}$ (2.24 $\times$ 10$^{15}$ cm$^{-2}$) \cite{43h} and Fe$_{78}$Ga$_{13}$B$_9$/Bi$_{85}$Sb$_{15}$  (5.03 $\times$ 10$^{15}$ cm$^{-2}$) \cite{45_a1} heterostructures. $G_{eff}$ values of the order of $\sim$ 10$^{15}$ cm$^{-2}$ have been reported in similar heavy metal/FM systems, such as 1.7 $\times$ 10$^{15}$ in Ni$_{80}$Fe$_{20}$/Ta, 1.2 $\times$ 10$^{15}$ in Ta/Co$_{40}$Fe$_{40}$B$_{20}$,  and 3.4 $\times$ 10$^{15}$ cm$^{-2}$ in Co$_2$FeAl/Ta \cite{45_a2}-\cite{45_a4}. The factors contributing to the high $G_{eff}$ value in our Ta/FeGa films are discussed in Section S6 of the Supporting Information. Unlike indirect measurements like FMR, we probed the FeGa layer directly to calculate $G_{\uparrow\downarrow}$ using TRMOKE, which provides a more reliable method for such investigation. The large value of $G_{eff}$ suggests a strong spin pumping into the Ta layer and that FeGa serves as a highly effective spin current source.     
   
\vspace{7pt}

\begin{figure}[ht!] 
\centerline{\includegraphics[scale=0.6]{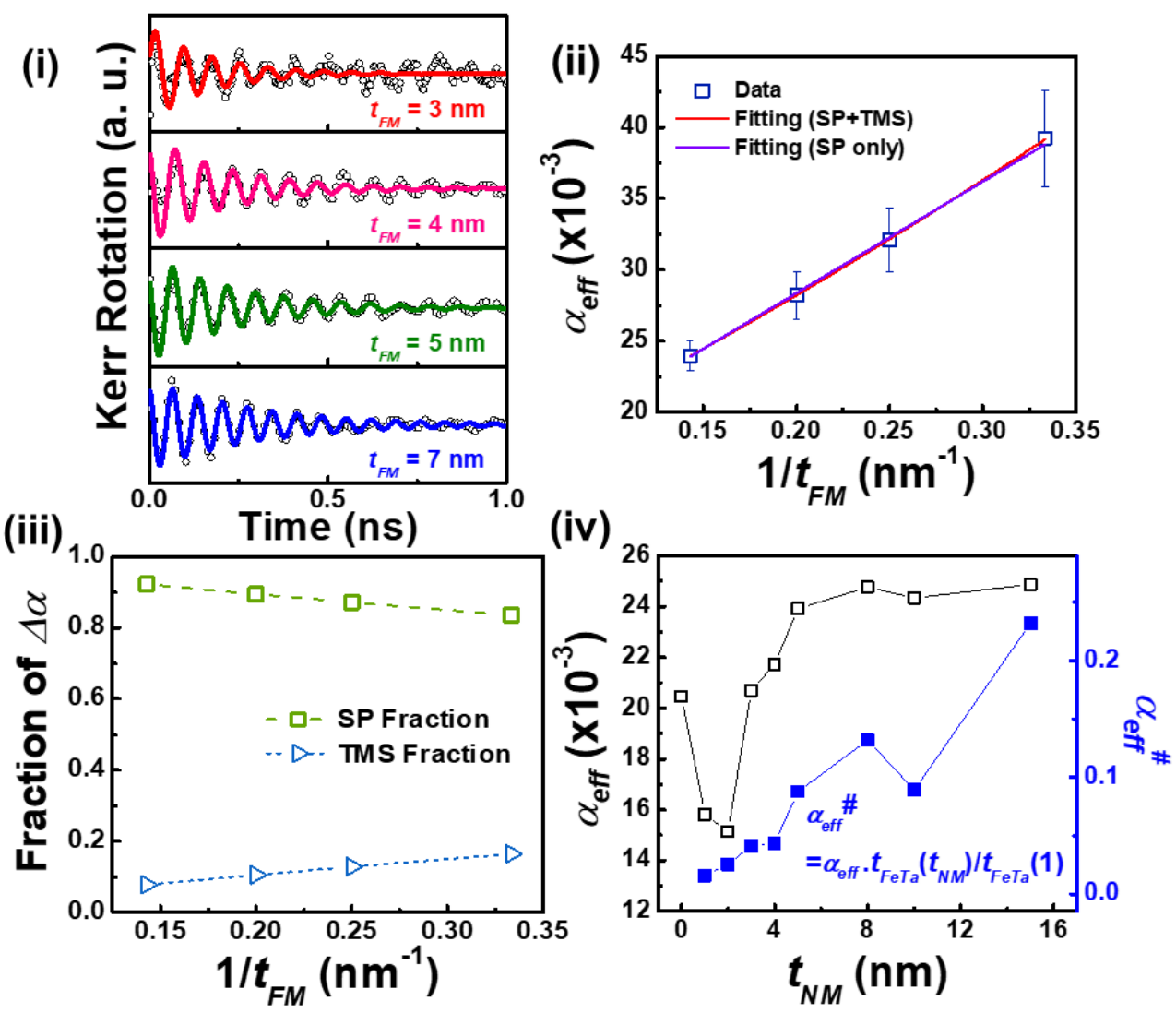}}
\caption{ Magnetization precession and extracted effective magnetic damping with FeGa thickness variation data are presented: (i) Background-subtracted precessional Kerr rotation data for Ta(5 nm)/FeGa ($t_{FM}$) films for $t_{FM}$ = 3, 4, 5, and 7 nm. (ii) Variation of $\alpha_{eff}$ with $1/t_{FM}$. Fitted curves considering only spin pumping (SP) as well as both SP and two-magnon scattering (TMS) are shown. (iii) Relative contributions of SP and TMS processes to the observed damping modulation $\Delta\alpha=(\alpha_{eff}-\alpha_0)$ (iv) Test of the role of spin memory loss (SML) arising from FeTa intermixing layer in the Gilbert damping modulation where $\alpha^{\#}_{eff}=\alpha_{eff}*[t_{FeTa}(t_{NM})⁄t_{FeTa}(t_{NM}=1 nm)]$.}\label{F6a}
\end{figure}

We note that $\alpha_{0}^{'}$ as extracted from fitting to Equation 9 includes a TMS component. To isolate the TMS effect from analysis, we can exploit the fact that the enhancement of Gilbert damping originating from the TMS effect shows a different functional dependence on the ferromagnetic thickness than that originating from the spin pumping. The $t_{FM}^{-2}$ dependence arises from the Arias-Mills model \cite{45_aa} where the TMS line-broadening is attributed to fluctuations in the interfacial surface anisotropy field due to the presence of magnetic inhomogeneities associated with defects at the interface. As a result, the spin pumping and TMS contributions to the observed damping enhancement can be straightforwardly isolated in an elegant manner by analyzing the ferromagnetic thickness-dependence of the Gilbert damping. Thus, to explicitly isolate the TMS component experimentally, as well as find the effective spin-mixing conductance $G_{eff}$, we performed additional TRMOKE measurements by varying the thickness of FeGa ($t_{FM}$=3,4,5 and 7 nm) for a fixed Ta UL thickness of $t_{NM}$ = 5 nm. The results are presented in Figure \ref{F6a}. The $t_{FM}$=4 nm trace could be better fit by superposing a higher frequency mode. However, we only consider the damping of the uniform Kittel mode for consistency throughout our analysis. We fit the $\alpha_{eff}$ versus $t_{FM}^{-1}$ data using Equation 8 in two scenarios, by either considering or neglecting the TMS term. When TMS is considered, the value of $\alpha_{0}$ is extracted to be 0.01429, $G_{eff}$ = (5.48 $\pm$ 0.034) $\times$ $10^{15}$ cm$^{-2}$ and $\beta_{TMS}$ = (3.68 $\pm$ 0.172) $\times$ $10^{-16}$ cm$^2$. Without considering TMS, an artificially higher value of $G_{eff}$ = (6.890 $\times$ $\pm$ 0.173) $\times$ $10^{15}$ cm$^{-2}$ and lower value of $\alpha_{0}$ = 0.01272 are extracted. Figure \ref{F6a} (iii) shows the relative contributions of the spin pumping and TMS effects to the damping modulation $\Delta\alpha=(\alpha_{eff}-\alpha_0)$. It can be seen from the figure that the TMS contribution to damping enhancement increases with decreasing FeGa thickness, but remains under 20\% within the experimentally studied thickness range. Thus, although a significant TMS component is present in the system, our experimental results are not quite in accord with the picture of dominating TMS contribution reported in certain works \cite{45_a}. 

\vspace{7pt}

In non-epitaxially grown NM/FM heterostructures, presence of some degree of interfacial intermixing as observed in our case (see Table S1 of the Supporting Information) can also result in spin memory loss (SML), in which the pumped spin current is dissipated at the interface itself, leading to an enhancement in Gilbert damping without enhancing spin current transmission to the NM.  This occurs due to spin-flip scattering at impurity and defect centers in the presence of interfacial roughness and intermixing, resulting in partial depolarization of the spin current at the interface \cite{45_b}. The SML is typically ruled out by control experiments using a Cu insertion layer between the NM and FM layer, as it can isolate SML without hindering spin pumping due to its long spin diffusion length \cite{43d}. However, the insertion of a Cu spacer can nontrivially modify the strain relaxation in FeGa owing to its magnetostrictive nature, which does not allow it to be used for conclusively isolating the SML effect \cite{45_c, 45_d}. The presence of a Cu UL has been found to cause large changes to the magnetic coercivity or soft magnetic property of FeGa \cite{45_c}. Thus, it is not feasible to use Ta/Cu/FeGa as a control system for comparison with Ta/FeGa, as the static magnetic characteristics will also differ in the two cases. Moreover, since the presence of a Cu UL also increases the compressive strain on FeGa as compared to Si substrate, the insertion of a Cu spacer can itself modify the strain relaxation in FeGa and result in large changes in its Gilbert damping. Due to these additional effects which can independently modulate the damping, we cannot use Cu spacer for conclusively isolating interfacial scattering effects such as SML from damping measurements (see Section S7, Supporting Information). Interfacial engineering by annealing can also independently modify the FMR of FeGa films by releasing compressive stresses \cite{45_d}, so it cannot be used to modulate the SML alone either. To investigate the role of SML, we have studied the correlation of the effective Gilbert damping with FeTa mixed layer thickness $t_{FeTa}$ which can contribute to the SML effect in our heterostructures. In Figure \ref{F6a} (iv), we present the variation of $\alpha_{eff}$ as well as the product $\alpha^{\#}_{eff}=\alpha_{eff}.[t_{FeTa}(t_{NM})⁄t_{FeTa}(t_{NM}=1 nm)]$, as a function of the Ta UL thickness $t_{NM}$. These results demonstrate that the increasing thickness of FeTa mixed layer does not lead to the observed $t_{NM}$-dependence of $\alpha_{eff}$. In other words, any SML effect due to FeTa mixed layer formation is not the major factor influencing the Gilbert damping modulation in our samples. 

\vspace{7pt}

However, we cannot entirely discount the possibility of a finite SML effect present in these heterostructures contributing non-trivially to the damping modulation. Since the SML contribution to the damping has a linear 1/$t_{FM}$ thickness dependence like the spin pumping \cite{45_a}, it cannot be ruled out by FM thickness-dependent measurements. Further, since in our case, using a weak spin scattering insertion layer of Cu to isolate SML is not possible without causing nontrivial modifications to the strain relaxation in magnetostrictive FeGa, the spin-mixing conductance values we have extracted for Ta/FeGa system are best interpreted as overall effective “upper-limit” values which can include a finite SML component. As discussed above, the magnetostriction of FeGa complicates the isolation of SML in heavy metal/FeGa systems via interface engineering, not only using spacer layers such as Cu, but also via annealing, as done in \cite{45_a}. Nonetheless, this route can be explored in more detail in a future study, where interfacial spin-orbit coupling (ISOC) strength at the Ta/FeGa interface is systematically tuned in several different series of thickness-varied sample sets, while measures are taken to isolate resulting modifications to strain relaxation in FeGa. Thereafter, the SML conductance can be indirectly derived by normalizing with the relative change of damping-like torque efficiency from its extrapolated value at zero ISOC strength \cite{45_a}. In addition to $G_{eff}$, the interfacial spin transparency ($T$) at the NM and FM interface is another important parameter for spintronics applications. Finite element analyses of spin-transport correlate $G_{eff}$ and $T$ at the Ta/FeGa interface via the thickness $t_{NM}$, spin diffusion length $\lambda_{sd}$, and resistivity $\rho$ of the Ta layer \cite{43d, 46, 47}. However, for calculating the interfacial transparency reliably, all other interfacial effects other than spin pumping (forward and backflow spin currents) must be explicitly excluded. Therefore, we have avoided calculating the interfacial spin transparency for our Ta/FeGa system, as it is difficult to interpret this quantity in the present case where SML cannot be excluded beyond doubt. Further studies are in order to identify appropriate candidate materials to apply as passivation spacers for isolating interfacial spin scattering effects in NM/FeGa systems without causing any appreciable changes to the strain relaxation in FeGa.

\vspace{7pt}
 
\begin{figure}[ht!] 
\centerline{\includegraphics[scale=0.7]{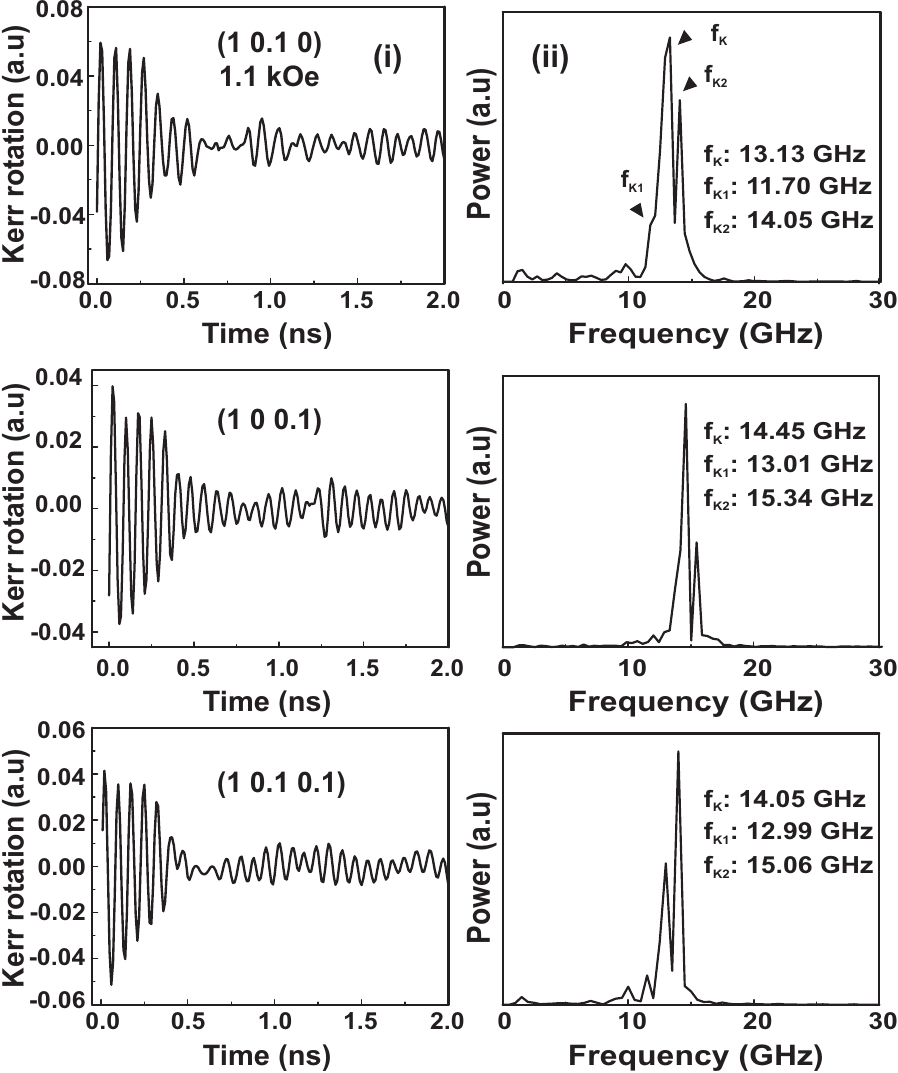}}
\caption{Simulated magnetization precession data and the corresponding fast Fourier transformed (FFT) power spectra are presented as follows: (i) Background-subtracted magnetization precession data for three different anisotropy configurations are shown at a field value 1.1 kOe, as mentioned in the inset. For the specific configurations mentioned: (1 0.1 0): this represents an anisotropy axis primarily along the x-direction with a small component, 10\% of the strength of the x-component, in the y-direction. (ii) The corresponding FFT power spectra are displayed, with $f_{K}$ representing the Kittel mode, while $f_{K1}$ and $f_{K2}$ correspond to additional modes, and their respective values are provided in inset of the graph.}\label{F7}
\end{figure} 
  
\vspace{7pt}

Finally, we performed micromagnetic simulations to understand the effect of tilted magnetic anisotropy in the presence of strain on the Kittel mode and other modes in FeGa films \cite{47}. The film dimensions are set at 1 $\mu$m $\times$ 1 $\mu$m $\times$ 8 nm, with an exchange stiffness constant of 2.1 $\times$ $10^{-17}$ erg/cm, a saturation magnetization of 1250 emu/cc, and an anisotropy constant with a value of -2.5 $\times$ $10^{5}$ ergcm$^{-3}$. These parameters have been optimized to closely match our experimental findings. It is worth noting that material parameters for bulk FeGa cannot be directly applied to FeGa thin films \cite{4}. Determining magnetostriction in a very thin layer, such as the 7 nm thick FeGa film in our case, can be challenging. We have used a reasonable magnetostriction value of 10 $\pm$ 5 ppm, taking following the reported values for FeGa films grown on Si substrates \cite{21}. In our simulations, we varied the orientation of magnetic anisotropy from IP to OOP direction. Figure \ref{F7}(i) illustrates the simulated magnetization precession ($m_Z$ component) at an applied field of 1.1 kOe for different anisotropy configurations (1 0.1 0), (1 0 0.1), and (1 0.1 0.1). The corresponding FFT power spectra shown in Figure \ref{F7}(ii). Here, $f_K$ corresponds to the Kittel uniform precession mode, while $f_{K1}$ and $f_{K2}$ represent other magnetic modes influenced by magnetostriction. The values of $f_{K1}$, $f_{K2}$, and $f_{K3}$ are presented in the inset of Figure \ref{F7}(ii). Notably, the value of $f_{K}$ = 14.05 GHz is close to the experimentally observed Kittel mode value, as indicated in Figure \ref{F5}(iii). Additionally, the values of $f_{K1}$ and $f_{K2}$ closely align with the experimentally observed values shown in Figure \ref{F5}(iv). Therefore, tilted magnetic anisotropy in FeGa films significantly impacts the emergence of the Kittel mode and other magnetic modes in the presence of magnetostriction.        

\section{Conclusion}\label{sec3}
We investigated the structural and static magnetic properties along with the ultrafast magnetization dynamics in Ta/FeGa thin film heterostructures with systematic variation of thickness of Ta underlayer (UL). These films show a bcc-phase and the formation of an interfacial layer is observed with increment of UL thickness. FeGa film exhibits soft magnetic properties, tilted magnetic anisotropy and irregular magnetic domains. Using time-resolved magneto-optical Kerr effect (TRMOKE) measurements, we observed Kittel mode and additional strain-induced modes in all the films. Our comprehensive analysis of the precessional decay time in FeGa films with Ta UL of various thicknesses contributes valuable insights to the existing literature on magnetization dynamics. The effective magnetic damping is observed at its lowest $\approx$ 0.015 in our case, and damping is enhanced as thickness of UL is increased. To account for the increment in damping in Ta/FeGa films, spin pumping and two-magnon scattering mechanism using a ballistic transport model are considered. FeGa thickness dependent measurements of precessional dynamics are carried out to determine the spin-mixing conductance and coefficient of two magnon scattering at the Ta/FeGa interface. Further, micromagnetic simulations are employed to verify the effect of tilted magnetic anisotropy and strain in the formation of magnetic modes. These findings in the FeGa material system offer new insights for controlling SW propagation and the exploration of diverse applications within the field of straintronics and can inform future research in the field. In particular, investigating spin-Hall and Rashba spin-orbit torques (SOTs) or study of the spin Hall angle as a function of Ta thickness through electrical or all-optical means in Ta/FeGa system can inspire future studies leading to important insights with regards to the interplay of interfacial lattice strain and spin transport in this technologically relevant system.

\section{Experimental Section}\label{sec4}
A series of Ta (0–15 nm) / FeGa (7 nm) thin films were deposited on bare Si substrates using RF magnetron sputtering at room temperature. Initially, a target with nominal composition $Fe_{73}Ga_{27}$ is considered. We used a 2 nm thick Ta capping layer for all the films, which oxidizes to form the native surface oxide and prevents oxidation of the underlying magnetic film. Ta capping layer protects FeGa and does not create any additional strain in it. We note that any residual spin relaxation occurring at the FeGa/Ta(top) interface will be present uniformly across all the samples studied, and will therefore not affect parameters extracted from Ta or FeGa thickness-dependent measurements. The sputtering power and pressure were kept constant at 150 W and 5 mTorr, respectively, for all the depositions. The base vacuum chamber pressure was better than 1 $\times$ $10^{-8}$ Torr prior to the film deposition. Low deposition rates of 0.38 \AA/s for Ta and 0.78-0.8 \AA/s for FeGa were used to ensure uniform deposition.

\vspace{7pt}

Structural characterization was performed using grazing incidence X-ray diffraction (GIXRD) and X-ray reflectivity (XRR) with source of Cu $K_{\alpha}$ radiation. Magnetic properties were measured using physical property measurement system (PPMS). The film composition was determined by employing energy dispersive X-ray spectroscopy (EDS) integrated with field emission scanning electron microscope (FESEM). For investigating the topographic features of the films, atomic force microscopy (AFM) was employed. Magnetic force microscopy (MFM) was utilized to probe the magnetic stray fields emanating from the film surface, while the measurement of angle-dependent coercive field and remnant magnetization was mapped using static magneto-optical Kerr effect (SMOKE), and vibrating sample magnetometer (VSM), respectively.

\vspace{7pt}

Time-resolved magnetization dynamics of films were probed using a time-resolved magneto-optical Kerr effect (TR-MOKE) magnetometer. We used two-color optical pump–probe experiment in a non-collinear geometry for all measurements. The output (known as fundamental beam) from a femtosecond amplifier laser system (LIBRA, Coherent; pulse width = 40 fs, repetition rate = 1 k$H_{z}$) with a wavelength of 800 nm was used as the probe beam, and the second harmonic of the fundamental beam with a wavelength of 400 nm was used as the pump beam. The pump and probe beams were focused to spot diameters of about 250 $\mu$m and 100  $\mu$m, respectively, on the sample surface. The pump beam perturbs the equilibrium state of the magnetization in the sample. Eventually, the effective magnetization vector precesses around the effective magnetic field direction. The probe beam is considered as weaker compared to the pump beam, and records the magnetic response after excitation by the pump. The probe beam was incident normal to the sample plane, and its back-reflection was collected through a beam splitter onto the detector assembly, which simultaneously measured the Kerr rotation and reflectivity signals as a function of the time delay between the pump and probe beams. A variable magnetic field was applied to the sample at a small tilt from the sample normal, the in-plane component of which is referred to as the bias magnetic field. The out-of-plane tilt ensures a finite demagnetizing field, which is modulated by the laser pump pulse to result in a change of the equilibrium magnetization orientation which triggers the initial precession of magnetization around its new orientation. The Kerr rotation and reflectivity signals were measured in a phase-sensitive manner using two separate lock-in amplifiers. Finally, the experimental findings were corroborated with micromagnetic simulations using the mumax$^3$ freeware.

\vspace{7pt}

\textbf{Acknowledgements} \par 

\vspace{7pt}

AKS would like to thank the Advanced Postdoctoral Research Programme at the S. N. Bose National Centre for Basic Sciences (SNBNCBS) for providing fellowship during this work. SM acknowledges Department of Science and Technology (DST), Govt. of India for the INSPIRE Fellowship. AB gratefully acknowledges financial assistances from the SNBNCBS, India under Project No. SNB/AB/11-12/96 and DST, Govt. of India (Grant No. DST/NM/TUE/QM-3/2019-1C-SNB). Authors would like to thank the Technical Research Centre at SNBNCBS for XRD, and PPMS characterizations. Authors acknowledge the DST - FIST facility, VSM at the Department of Physics, IIT Hyderabad (Project No.: SR/FST/PSI-215/2016).   



\begin{thebibliography}{26}

\bibitem{1}
 A. E. Clark, J. B. Restorff, M. Wun-Fogle, T. A. Lograsso, D. L. Schlagel, \textit{IEEE Trans. Magn.} \textbf{2000}, \textit{36}, 3238.

\bibitem{2} 
G. Petculescu, K. B. Hathaway, T. A. Lograsso,  M. Wun-Fogle,  A. E. Clark, \textit{ J. Appl. Phys.} \textbf{2005}, \textit{97}, 10.

\bibitem{3}
N. Srisukhumbowornchai, S. Guruswamy, \textit{ J. Appl. Phys.} \textbf{2001}, \textit{90}, 5680.

\bibitem{4} 
J. Atulasimha, A. B. Flatau, \textit{Smart Mater. Struct.} \textbf{2011}, \textit{20}, 043001.

\bibitem{5}
 J. A. Chelvane, A. Talapatra, J. Mohanty, \textit{Mater. Res. Express} \textbf{2019}, \textit{6}, 116120.

\bibitem{6}
W. Jahjah, R. Manach, Y. Le Grand, A. Fessant, B. Warot-Fonrose, A. R. E. Prinsloo, C. J. Sheppard, D. T. Dekadjevi, D. Spenato, J. Ph. Jay, \textit{Phys. Rev. Appl.} \textbf{2019}, \textit{12}, 024020.

\bibitem{7}
 E. D. T. de Lacheisserie, Magnetostriction: Theory and Applications of Magnetoelasticity, \textit{CRC Press} \textbf{1993}.

\bibitem{8}  
G. Engdahl, I. D. Mayergoyz, Handbook of giant magnetostrictive materials, \textit{San Diego: Academic press} \textbf{2000}, 386.

\bibitem{9} 
W. Li, B. Buford, A. Jander, P. Dhagat, \textit{IEEE Trans. Magn.} \textbf{2014}, \textit{50}, 37.

\bibitem{10} 
J. Sun, B. Zhang, H. E. Katz, \textit{Adv. Funct. Mater.} \textbf{2011}, \textit{21}, 29.

\bibitem{10a}
A. Javed, N.A. Morley, M.R.J. Gibbs, \textit{J. Magn. Magn. Mater.} \textbf{2009}, \textit{321}, 2877.

\bibitem{10b}
C. Dong, M. Li, X. Liang, H. Chen, H. Zhou, X. Wang, Y. Gao, M. E. McConney, J. G. Jones, G. J. Brown, B. M. Howe, N. X. Sun, \textit{Appl. Phys. Lett.} \textbf{2018}, \textit{113}, 262401. 

\bibitem{10c}
Y. Wang, K. Yadagiri, P. Wu, T. Wu, \textit{AIP Advances} \textbf{2022} \textit{12}, 035027.

\bibitem{10d}
K. Yadagiri, Y. Wang, T. Wu, \textit{Mater. Chem. Phys.} \textbf{2022}, \textit{279}, 125776.

\bibitem{11}  
H. Ahmad, J. Atulasimha, S. Bandyopadhyay, \textit{Sci. Rep.} \textbf{2015}, \textit{1}, 18264.

\bibitem{12}
 A. Acosta, K. Fitzell, J. D. Schneider, C. Dong, Z. Yao, Y. E. Wang, G. P. Carman, N. X. Sun, J. P. Chang, \textit{Appl. Phys. Lett.} \textbf{2020}, \textit{116}.

\bibitem{13}
 A. E. Clark, K. B. Hathaway, M. Wun-Fogle, J. B. Restorff, T.  A. Lograsso, V. M. Keppens, G. Petculescu, R. A. Taylor,  \textit{J. Appl. Phys.} \textbf{2003}, \textit{93}, 8621.

\bibitem{14} 
 A. Emdadi, V. V. Palacheva, V. V. Cheverikin, S. Divinski, G. Wilde, I. S. Golovin, \textit{J. Alloys Compd.} \textbf{2018}, \textit{758}, 214.

\bibitem{15} 
G. A. Ramírez,  A. Moya-Riffo, D. Goijman, J. E. Gomez, F. Malamud, L. M. Rodríguez, D. Fregenal, A. Butera, J. Milano, \textit{J. Magn. Magn. Mater.} \textbf{2021},  \textit{535}, 168047. 

\bibitem{15-a} 
P. Bartolom\'e, M. Maicas, R. Ranchal, \textit{J. Magn. Magn. Mater.} \textbf{2020}, \textit{514}, 167183.

\bibitem{16} 
T. Brintlinger, S. H. Lim, K. H. Baloch, P. Alexander, Y. Qi, J. Barry, J. Melngailis, L. Salamanca-Riba, I. Takeuchi, J. Cumings, \textit{Nano Lett.} \textbf{2010}, \textit{10}, 1219.

\bibitem{17} 
 A. K. Sahoo, J. A. Chelvane, J. Mohanty, \textit{Appl. Phys. A} \textbf{2023}, \textit{129}, 419.

\bibitem{18} 
 A. Barman, J. Sinha, Spin dynamics and damping in ferromagnetic thin films and nanostructures, \textit{Springer International Publishing}, Cham \textbf{2018}.

\bibitem{19} 
A. Okada, S. He, B. Gu, S. Kanai, A. Soumyanarayanan, S. T. Lim, M. Tran, M. Mori, S. Maekawa, F. Matsukura, H. Ohno, \textit{Proc. Natl. Acad. Sci.} \textbf{2017}, \textit{114}, 3815.

\bibitem{20} 
A. A. Rzhevsky, B. B. Krichevtsov, D. E. B\"urgler, C. M. Schneider, \textit{Phys. Rev. B} \textbf{2007}, \textit{75}, 224434.

\bibitem{21} 
A. Acosta, K. Fitzell, J. D. Schneider, C. Dong, Z. Yao, R. Sheil, Y. E. Wang, G. P. Carman, N. X. Sun, J. P. Chang, \textit{J. Appl. Phys.} \textbf{2020}, \textit{128}.

\bibitem{22} 
S. N. Panda, S. Mondal, J. Sinha, S. Choudhury, A. Barman, \textit{Sci. Adv.} \textbf{2019}, \textit{5}, eaav7200. 

\bibitem{22-a}
F. Hellman, A. Hoffmann, Y. Tserkovnyak, G. S. D. Beach, E. E. Fullerton, C. Leighton, A. H. MacDonald, D. C. Ralph, D. A. Arena, H. A. Durr, P. Fischer, J. Grollier, J. P. Heremans, T. Jungwirth, A. V. Kimel, B. Koopmans, I. N. Krivorotov, S. J. May, A. K. Petford-Long, J. M. Rondinelli, N. Samarth, I. K. Schuller, A. N. Slavin, M. D. Stiles, O. Tchernyshyov, A. Thiaville, B. L. Zink, \textit{Rev. Mod. Phys.} \textbf{2017}, \textit{89}, 025006.

\bibitem{22-b}
J. Sinova, S. O. Valenzuela, J. Wunderlich, C. H. Back, T. Jungwirth, \textit{Rev. Mod. Phys.} \textbf{2015}, \textit{87}, 1213.

\bibitem{22c}
K. Uchida, S. Takahashi, K. Harii, J. Ieda, W. Koshibae, K. Ando, S. Maekawa, E. Saitoh, \textit{Nature} \textbf{2008}, \textit{455}, 778.

\bibitem{22d}
W. Lin, M. Hehn, L. Chaput, B. Negulescu, S. Andrieu, F. Montaigne, S. Mangin, \textit{Nat. Commun.} \textbf{2012}, \textit{3}, 744.

\bibitem{22e}
Y. Tserkovnyak, A. Brataas, G. E. Bauer, \textit{Phys. Rev. Lett.} \textbf{2002}, \textit{88}, 117601.

\bibitem{22f}
V. E. Demidov, S. Urazhdin, R. Liu, B. Divinskiy, A. Telegin, S. O. Demokritov, \textit{Nat. Commun.} \textbf{2016}, \textit{7}, 10446

\bibitem{23}  
A. Kumar, S. Akansel, H. Stopfel, M. Fazlali, J. Åkerman, R. Brucas, P. Svedlindh, \textit{Phys. Rev. B} \textbf{2017}, \textit{95}, 064406.

\bibitem{24}   
A. Brataas, Y. Tserkovnyak, G. E. W. Bauer, P. J. Kelly, \textit{Spin current (Oxford University Press)} \textbf{2012}, \textit{17}, 87.

\bibitem{25}  
S. Mukhopadhyay, P. K. Pal, S. Manna, C. Mitra, A. Barman, \textit{Phys. Rev. B} \textbf{2024}, \textit{109}, 024437.

\bibitem{26}  
R. Cheng, J. Xiao, Q. Niu, A. Brataas, \textit{Phys. Rev. Lett.} \textbf{2014}, \textit{113}, 057601.

\bibitem{27}  
K. Ando, S. Takahashi, J. Ieda, Y. Kajiwara, H. Nakayama, T. Yoshino, K. Harii, Y. Fujikawa, M. Matsuo, S. Maekawa, E. Saitoh, \textit{J. Appl. Phys.} \textbf{2011}, \textit{109}.

\bibitem{28} 
K. Ando, Y. Kajiwara, S. Takahashi, S. Maekawa, K. Takemoto, M. Takatsu, E. Saitoh, \textit{Phys. Rev. B} \textbf{2008}, \textit{78}, 014413.

\bibitem{29}  
M. Weiler, J. M. Shaw, H. T. Nembach, T. J. Silva, \textit{Phys. Rev. Lett.} \textbf{2014}, \textit{113}, 157204.

\bibitem{30}  
D. B. Gopman, V. Sampath, H. Ahmad, S. Bandyopadhyay, J. Atulasimha, \textit{IEEE Trans. Magn.} \textbf{2017}, \textit{53}, 1.

\bibitem{31}  
W. Zhang, Q. Liu, Z. Yuan, K. Xia, W. He, Q. Zhan, X. Zhang, Z. Cheng, \textit{Phys. Rev. B} \textbf{2019}, \textit{100}, 104412.

\bibitem{32}  
Y. Zhang, C. Huang, Z. Duan, W. Shi, \textit{J. Magn. Magn. Mater.} \textbf{2019}, \textit{477}, 4.

\bibitem{33}   
W. K. Peria, X. Wang, H. Yu, S. Lee, I. Takeuchi, P.  A. Crowell, \textit{Phys. Rev. B} \textbf{2021}, \textit{103}, L220403.

\bibitem{34}
B. D. Cullity, Elements of X-ray Diffraction, \textit{Addison-Wesley Publishing} \textbf{1956}.

\bibitem{34-a}
M. Björck, G. Andersson, \textit{J. Appl. Crystallogr.} \textbf{2007}, \textit{40}, 1174.

\bibitem{35}
D. Cao, X. Cheng, L. Pan, H. Feng, C. Zhao, Z. Zhu, Q. Li, J. Xu, S. Li, Q. Liu, J. Wang, \textit{AIP Adv.} \textbf{2017}, \textit{7}. 

\bibitem{37}
A. Begué, M. G. Proietti, J. I. Arnaudas, M. Ciria, \textit{J. Magn. Magn. Mater.} \textbf{2020}, \textit{498}, 166135.

\bibitem{38}
Z. Zhang, E. Liu, X. Lu, W. Zhang, Y. You, G. Xu, Z. Xu, P. K. J. Wong, Y. Wang, B. Liu, X. Yu, \textit{Adv. Funct. Mater.} \textbf{2021}, \textit{31}, 2007211.  

\bibitem{38a}
S. Mondal, M. A. Abeed, K. Dutta, A. De, S. Sahoo, A. Barman, S. Bandyopadhyay, \textit{ACS Appl. Mater. Interfaces} \textbf{2018}, \textit{10}, 43970.

\bibitem{39}
L. Sheng, Y. Liu, J. Chen, H. Wang, J. Zhang, M. Chen, J. Ma, C. Liu, S. Tu, C. W. Nan, H. Yu, \textit{Appl. Phys. Lett.} \textbf{2020}, \textit{117}. 

\bibitem{40}
B. H. Lee, T. Fakhrul, C. A. Ross, G. S. Beach, \textit{Phys. Rev. Lett.} \textbf{2023}, \textit{130}, 126703.

\bibitem{41}
R. Urban, G. Woltersdorf, B. Heinrich, \textit{Phys. Rev. Lett.} \textbf{2001}, \textit{87}, 217204.

\bibitem{42}
S. Rafique, J. R. Cullen, M. Wuttig, J. Cui, \textit{J. Appl. Phys.} \textbf{2004}, \textit{95}, 6939.   

\bibitem{43}
T. Nan, S. Emori, C. T. Boone, X. Wang, T. M. Oxholm, J. G. Jones, B. M. Howe, G. J. Brown, N. X. Sun, \textit{Phys. Rev. B} \textbf{2015}, \textit{91}, 214416.

\bibitem{43aa}
J. E. Hirsch, \textit{Phys. Rev. Lett.} \textbf{1999}, \textit{83}, 1834.

\bibitem{43a}
S. Emori, A. Matyushov, H. -M. Jeon, C. J. Babroski, T. Nan, A. M. Belkessam, J. G. Jones, M. E. McConney, G. J. Brown, B. M. Howe, N. X. Sun, \textit{Appl. Phys. Lett.} \textbf{2018}, \textit{112}, 182406.

\bibitem{43b}
P. M. Haney, H. -W. Lee, K. -J. Lee, A. Manchon, M. D. Stiles, \textit{Phys. Rev. B} \textbf{2013}, \textit{87}, 174411.

\bibitem{43c}
Q. Liu, Y. Zhang, L. Sun, B. Miao, X. R. Wang, H. F. Ding, \textit{Appl. Phys. Lett.} \textbf{2021}, \textit{118}, 132401.

\bibitem{43d}
W. Zhang, W. Han, X. Jiang, S. H. Yang, S. S. P. Parkin, \textit{Nat. Phys.} \textbf{2015}, \textit{11}, 496.

\bibitem{43e}
Y. Du, S. Takahashi, J. Nitta, \textit{Phys. Rev. B} \textbf{2021}, \textit{103}, 094419.

\bibitem{43f}
E. Saitoh; M. Ueda; H. Miyajima; G. Tatara, \textit{Appl. Phys. Lett.} \textbf{2006}, \textit{88}, 182509.

\bibitem{43g}
E. Longo, M. Belli, M. Alia, M. Rimoldi, R. Cecchini, M. Longo, C. Wiemer, L. Locatelli, P. Tsipas, A. Dimoulas, G. Gubbiotti, M. Fanciulli, R. Mantovan, \textit{Adv. Funct. Mater.} \textbf{2022}, \textit{32}, 2109361.

\bibitem{43h}
V. Sharma, P. Bajracharya, A. Johnson, and R. C. Budhani, \textit{AIP Adv.} \textbf{2022}, \textit{12}, 035028.

\bibitem{44}
Y. Tserkovnyak, A. Brataas, G. E. Bauer, \textit{Phys. Rev. B} \textbf{2002}, \textit{66}, 224403.

\bibitem{44b}
S. Mukhopadhyay, P. K. Pal, S. Manna, C. Mitra, A. Barman, \textit{NPG Asia Mater.} \textbf{2023}, \textit{15}, 57.

\bibitem{45}
J. M. Shaw, H. T. Nembach, T. J. Silva,  \textit{Phys. Rev. B} \textbf{2012}, \textit{85}, 054412. 

\bibitem{45_a} 
L. Zhu, D. C. Ralph, R. A. Buhrman, \textit{Phys. Rev. Lett.} \textbf{2022}, \textit{5}, 057203.   

\bibitem{45_a1} 
V. Sharma, W. Wu, P. Bajracharya, D. Q. To, A. Johnson, A. Janotti, G. W. Bryant, L. Gundlach, M. B. Jungfleisch, and R. C. Budhani, \textit{Phys. Rev. Mater.} \textbf{2021}, \textit{5}, 124410.

\bibitem{45_a2} 
E. Montoya, P. Omelchenko, C. Coutts, N. R. Lee-Hone, R. Hübner, D. Broun, B. Heinrich, E. Girt, \textit{Phys. Rev. B} \textbf{2016}, \textit{94}, 054416. 

\bibitem{45_a3} 
R. Yu, B. F. Miao, L. Sun, Q. Liu, J. Du, P. Omelchenko, B. Heinrich, Mingzhong Wu, H. F. Ding, \textit{Phys. Rev. Mater.} \textbf{2018}, \textit{2}, 074406. 

\bibitem{45_a4} 
A. Kumar, R. Gupta, S. Husain, N. Behera, S. Hait, S. Chaudhary, R. Brucas, P. Svedlindh, \textit{Phys. Rev. B} \textbf{2019}, \textit{100}, 214433. 

\bibitem{45_aa}
R. Arias and D. L. Mills, \textit{Phys. Rev. B} \textbf{1999}, \textit{60}, 7395.

\bibitem{45_b}
J.-C. Rojas-Sánchez, N. Reyren, P. Laczkowski, W. Savero, J.-P. Attané, C. Deranlot, M. Jamet, J.-M. George, L. Vila, and H. Jaffrès, \textit{Phys. Rev. Lett.} \textbf{2014}, \textit{112}, 106602.

\bibitem{45_c}
A. Acosta, K. Fitzell, J. D. Schneider, C. Dong, Z. Yao, Y. E. Wang, G. P. Carman, N. X. Sun, J. P. Chang, \textit{Appl. Phys. Lett.} \textbf{2020}, \textit{116}, 222404.

\bibitem{45_d}
X. Guo, F. Wang, X. Ma, Q. Li, M. Liu, X. Chen, J. Yu, J. Xu, S. Li, J. Wang, Q. Liu, D. Cao, \textit{Appl. Phys. Lett.} \textbf{2022}, \textit{120}, 202402.

\bibitem{46}
A. Brataas, Y. V. Nazarov, G. E. W. Bauer, \textit{Phys. Rev. Lett.} \textbf{2000}, \textit{84}, 2481. 

\bibitem{47}
A. Vansteenkiste, J. Leliaert, M. Dvornik, M. Helsen, F. Garcia-Sanchez, B. Van Waeyenberge, \textit{AIP Adv.} \textbf{2014}, \textit{4}, 107133. 

\end{thebibliography}
\end{document}